\documentclass[useAMS,usenatbib,useepsfig]{mn2e}

\usepackage{epsfig}
\usepackage{amssymb}
\usepackage{epstopdf}

\title[DGL at high $|b|$]
{Diffuse Galactic light at high Galactic latitude: nature and interpretation}
\author[Zagury, Fr\'ed\'eric]
{Fr\'ed\'eric Zagury \thanks{e-mail:fzagury@wanadoo.fr}\\
Institut Louis de Broglie, 23 rue Marsoulan, 75012 Paris, France}

\begin{document}

\date{Received 2006 March}

\pagerange{\pageref{firstpage}--\pageref{lastpage}} \pubyear{2002}

\maketitle

\label{firstpage}

\begin{abstract}
The hypothesis of an extended red emission (ERE) in diffuse Galactic light (DGL) has been put forward in 1998 by Gordon, Witt \& Friedmann who found that scattered starlight was not enough to explain the amount of DGL in the R band, in some high Galactic  latitude directions.
This paper re-investigates, for high Galactic latitudes, the brightnesses and colors of DGL, integrated starlight (ISGL), and of the total extra-solar light (ISGL+DGL) measured by Pioneer. 
Under the traditional assumption that DGL is forward scattering of background starlight by interstellar dust on the line of sight,  ISGL and Pioneer have very close colors, as it is found in Gordon, Witt \& Friedmann (1998).
Pioneer observations at high $|b|$ thus accept an alternative and simple interpretation, with no involvement of ERE in DGL.
\end{abstract}


    \section{Introduction}
Existence of a diffuse Galactic light component (DGL) in the night sky, at  visible wavelengths, arising from the scattering of starlight by interstellar matter in the Galaxy, was established in the 1930s \citep{struve36, elvey37}. 
It was quickly realized that the scattering is not of Rayleigh type, but that most of the scattered starlight is thrown in the direction of the incident beam; and that the grains have a large albedo  \citep{struve37, henyey41}. 
The particles responsible for the scattering  must therefore be large interstellar grains, rather than fine cosmic dust, presumably the same which are responsible for the linear visible extinction  \citep{stebbins39}.
Interstellar extinction and DGL occur  in discrete, structured, Galactic clouds, in which dust grains and gas are intimately mixed  \citep{lilley55}, and not from a pervasive, homogenous medium [\citet{schalen29} and \citet{stebbins39} for extinction;  \citet{mattila70, roach72} for DGL].

In this paper I shall be interested in DGL at high Galactic latitudes ($|b|\geq 30^\circ$), which is much weaker than towards the Milky Way (fig.~4 in \citet{roach72}), but may be easier to interpret because of the low column densities generally found along these sight-lines.
Early observation of DGL at high $|b|$ can be found in \citet{devau55} (see \citet{zagury00}), on Palomar plates \citep{lynds65}, and in \citet{sandage76}.
In the late 1970s' and in the 1980s', two major advances in astrophysical means of observation 
completed our understanding of DGL, and allowed a better estimate of its strength  at high Galatic latitudes.
Firstly, night sky brightness surveys have been done from 
outside our atmosphere (Pioneer observations, \citet{toller81,toller87}), thus facilitating evaluation of extra 
solar system light (DGL + 
direct starlight), especially at the high $|b|$, since atmospheric effects are 
removed.

\begin{figure}
 \resizebox{\columnwidth }{!}{\includegraphics{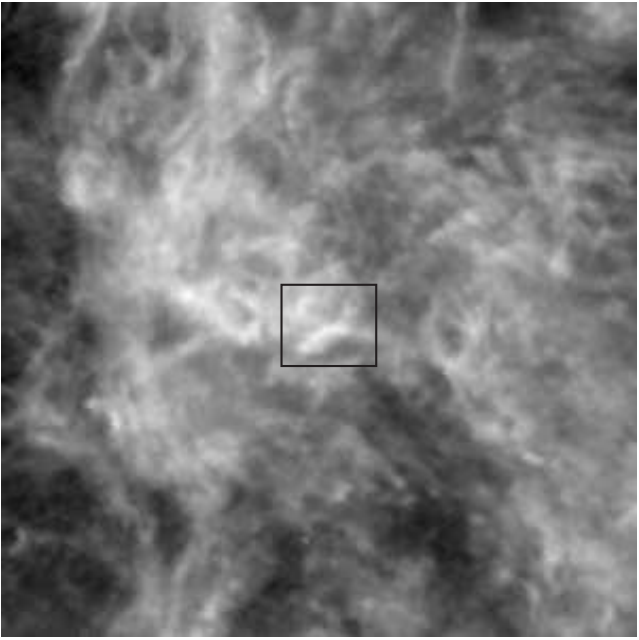}} 
\caption{IRAS $100\,\mu$m image (equatorial coordinates) centered on Polaris. The field is 
$13^{\circ}\times 13^{\circ}$ wide.} 
\label{fig:fig1}
\end{figure}

\begin{figure}
\resizebox{\columnwidth }{!}{\includegraphics{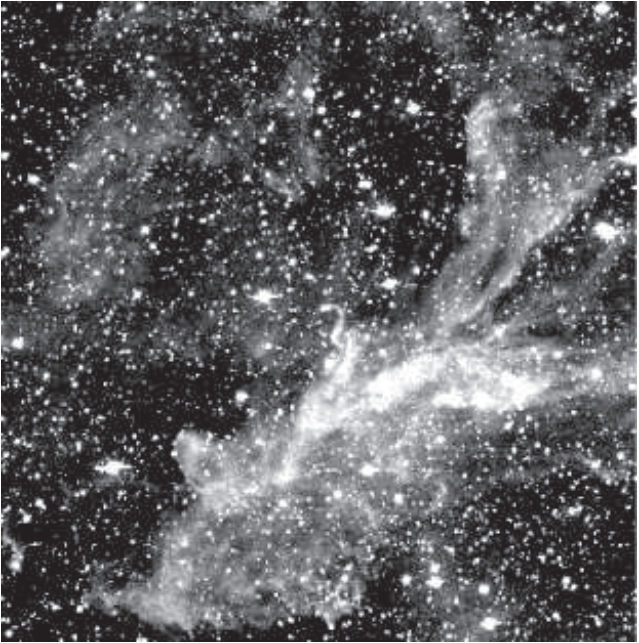}} 
\caption{I-band image of the region limited by the white square
in Fig.~\ref{fig:fig1}, magnified 150 
times (resolution = 2''). The field is $\sim 50'\times 50'$ wide, centered 
on $\alpha_{1950}¥=01h33'\,\beta_{1950}¥=87^{\circ}¥38'$. 
The image shows that the structure of
interstellar cirrus -still not resolved in this image- extends to much 
smaller scales than observed by IRAS. 
Scattering by forward scattering grains fully accounts for the visible 
brightness of the cloud \citep{zagury99} ruling out any emission process.
} 
\label{fig:fig2}
\end{figure}

Secondly, the InfraRed Astronomical Satellite (IRAS)  $100\,\rm\mu m$ images have revealed the thermal emission 
of the same interstellar large grains responsible for the visible interstellar extinction and for DGL, and give a direct view to the interstellar 
clouds in which DGL takes its origin.
IRAS shows that  the high latitude sky is largely covered by cirrus-like interstellar clouds with average low column densities  (Figs.~\ref{fig:fig1}, \ref{fig:fig2}, \ref{fig:fig3}).
These high latitude clouds (HLCs) appear as extended layers of dust and gas, with filamentary structures, and represent a natural 'barrier' between the background starlight and us.
They usually lie within 1~kpc from the sun.
Following the interpretation of DGL recalled above, DGL from one high latitude direction should be nothing but the light from background 
stars close to this direction, forward-scattered by interstellar grains within the interstellar cirrus on the line of sight.
The relative proximity of HLCs from the sun is  a non-negligible parameter for the proportion of DGL in Pioneer observations: the farther they are, the less scattered light we will receive because a smaller proportion of stars will participate in their illumination (the dilution of scattered starlight with distance should also intervene, but to a lesser extent, since most of it is thrown in the forward direction).
High latitude nebulosities discussed in \citet{sandage76} are HLCs' regions of local brightness enhancements,  either because of a local increase of the incident radiation field (due to the proximity of a bright star, for instance), a closer distance of the HLC to the sun, or because of higher column densities than the average. 

Observation of HLCs in the visible will generally require specific methods, like the recent CCD imaging techniques of low brightness objects, introduced by \citet{gu89}.
Even so, the visible images one obtains are confusing most of the time (figs.~2, 3, and 4 in \citet{gu89}).
Only one field out of the program initiated by \citet{gu89},  MCLD~123.5~+24.9 in the Polaris Flare ($\sim 
100$~pc from the sun), gave detailed high resolution images, in the B, R and I bands, of the small scale structure 
a high latitude cloud can have, at a resolution of $\sim 2''$ (150 times better than IRAS).
These images (Fig.~\ref{fig:fig2}) zoom into an HLC close to Polaris, and prove that interstellar cirrus are structured to 
much smaller scales than previously recorded by IRAS.
Their analysis and comparison with IRAS 
$100\,\rm\mu m$ image of the field (Zagury, Boulanger \& Banchet 1999), demonstrate that the brightnesses 
of the cloud in the R and I bands is due to forward scattering by large 
grains, in agreement with the interpretation of DGL.
It further indicates that the observed region consists in a widespread medium of low optical depth (which fills the entire image of Fig.~\ref{fig:fig2}), within which structures of much higher column densities (the spectacular features on the figure) are observed.

Successive improvements in the observation of DGL have permitted a refined reflection on its  nature, the main question being to know whether or not forward scattering by large interstellar grains explains all of DGL.
Isotropic scattering by the gas or by tiny dust, or a backscattering lobe in the scattering cross-section of the large grains, for instance, could contribute  to DGL.
Emission processes could also participate in some wavelength range.

An isotropic phase function, or a strong back-scattering lobe, would mean that DGL is, at least in part, reflected light.
This has been suggested by \citet{witt68} after an analysis of observations in the same areas as \citet{henyey41}, or by \citet{sandage76}, whose calculations showed that high latitude nebulosities observed in the visible could be reflected light from the Galactic plane (an hypothesis suggested by \citet{bergh66}).
None of these hypothesis however seem to hold.
\citet{vandehulst69}  (see also \citet{mattila71}), proved that a rigorous treatment of the radiative transfer used by \citet{witt68}, applied to the same observations, leads to a strong forward scattering phase function and a large albedo, as \citet{henyey41} had found.
The van der Bergh~-~SandageÕ's proposal that HLCs are illuminated by the Galactic plane is also problematic \citep{zagury00}, since dense clumps at high latitude are observed in absorption. 
The findings of \citet{toller81}, that the ratio of DGL  to direct integrated starlight (ISGL) decreases with increasing latitude and has no pronounced dependence on longitude, would further be difficult to understand, since, if illumination comes from the Galactic plane, DGL in one direction would not depend on ISGL in the same direction (and should vary with longitude).
Therefore, there is at present no indication that backscattering plays a significant role in DGL, a result I will use in the rest of the paper. 

The possibility of an emission process in DGL at high $|b|$, in addition to scattering,  was recently advanced by \citet{gordon98} (GWF hereafter), who find that some $50\% $ of 
DGL, in the red, comes
from luminescence, an emission process (ERE for `Extended Red 
Emission') of some, yet still unknown 
particles, which would re-emit in the red the energy they absorb at 
shorter wavelengths.
This discovery was strengthened by the \citet{gu89} (see also \citet{gu94}) article, who found ERE at high $|b|$ from an analysis of  visible B and R images of HLCs.

The  \citet{gu89}  detection of ERE in HLCs is based on SandageÕ's hypothesis of an illumination by the Galactic plane, and on an isotropic phase function, which, as recalled above, are not verified by observation.
Furthermore, the red light from MCLD~123.5~+24.9 is well explained by scattering alone, excluding any emission process :
if, in MCLD~123.5~+24.9, red filaments over the field are red, it is because they are denser, 
and therefore extinguish blue light more quickly than red light.
In the densest cores (as the dense clump detected in CO by \citet{falgarone98}, and seen in absorption on the visible images), 
all B and R light is extinguished, leaving only small residual 
brightness in the I band, in conformity with what is expected from 
scattering.

The finding of ERE at high Galactic latitude  meets other important difficulties.
Which are these particles that 
must be numerous enough to produce such a large emission in the red, and will require an extreme luminescence photon conversion efficiency close to $100\,\%$ \citep{zubko99}?
Why can't the phenomenon be reproduced on earth where we 
have at our disposal a much more
sophisticated chemistry than can be presumed to be the case in the 
basic and very low density conditions of an average high latitude 
cirrus?

There are, therefore, reasons to question the GWF finding of ERE at high $|b|$, and to confront the GWF data with what one can expect from forward scattering by large grains in HLCs.
This  is a critical issue since ERE has become an important and unresolved topic in the study of the interstellar medium.

This paper will be divided in two.
The first part,  Sects.~\ref{not} to \ref{obs}, is an analysis of the GWF data and of the detection of ERE in DGL.
In the second part,  the low average column density of HLCs will be used to propose an approach for the calculation of HLCs' brightness which involves only a few assumptions (a strong forward scattering phase function and a large albedo, as it has been justified previously).
Orders of magnitude will be calculated for the brightnesses and colors of ISGL, DGL, and Pioneer, in the direction of a low column density cirrus, assuming DGL is scattering alone.
An important outcome will be to verify whether or not these estimates can explain the very close colors found for Pioneer and integrated starlight in GWF.
 \section{Notations and data} \label{not}
I follow GWF's notations which define, for a given direction, in units 
of  intensities per 
unit solid angle:
$P_B$ and $P_R$, the Pioneer surface brightnesses in the B 
and R bands, after zodiacal light has been substracted; 
$isgl_B$ and $isgl_R$, the `integrated star/Galaxy light' (ISGL) for 
objects with V-magnitudes larger than 6.5 ($m_V>6.5$), 
estimated from a compilation of several star catalogues;
$S_{B}$ and $S_{R}$ the B and R brightnesses of 
DGL.
Stars brighter than $6.5$~mag in V have been substracted to the data for the calculation of these quantities.

With these definitions
\begin{eqnarray}
P_{R}\, = \, S_{R} +isgl_{R} \nonumber\\
P_{B}\, = \, S_{B} +isgl_{B}
\label{eq:p}
\end{eqnarray}¥
$P_B$ and $P_R$, $isgl_B$ and $isgl_R$, and $S_B$ and 
$S_R$, are estimated in boxes $5^{\circ}\times 5^{\circ}$ large, and  
along two cuts at Galactic longitudes $l=0$, between latitudes 
$b=30^{\circ}¥$ 
and $b=60^{\circ}¥$, and $l=100$, with $-55^{\circ}¥<b<-20^{\circ}¥$, called respectively region (a)  and 
region (b).

\begin{figure}
\resizebox{\columnwidth }{!}{\includegraphics{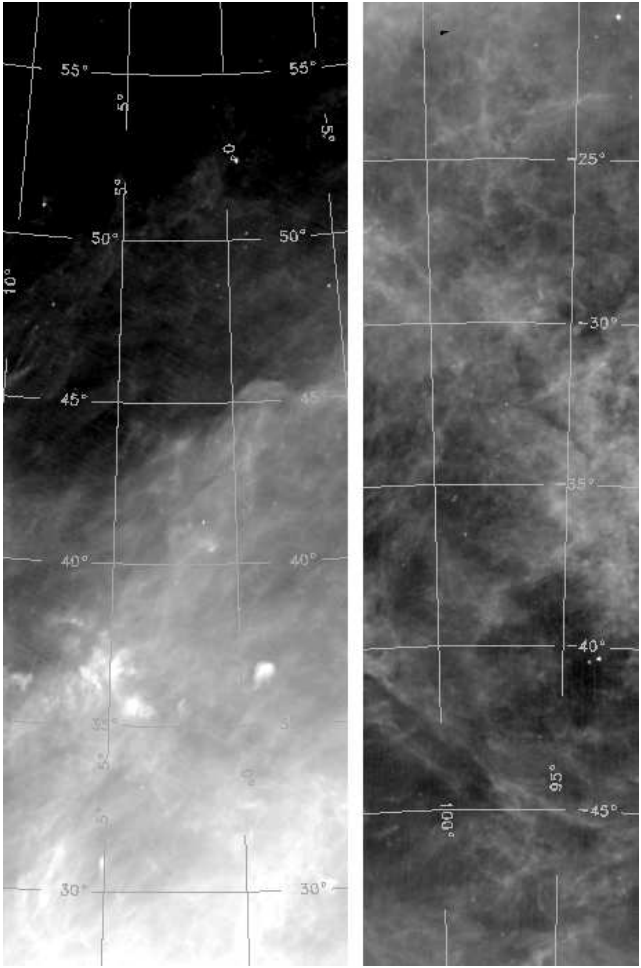}} 
\caption{IRAS $100\,\mu$m images of GWF fields (a), left, and (b), 
right. The brightest regions (saturated) of field (a) have surface 
brightnesses between $25$ and $30\,$MJy/sr. The surface 
brightness decreases to $11\,$MJy/sr at $40^{\circ}$. It is 
respectively $\sim3\,$MJy/sr and $\sim 1\,$MJy/sr in the brightest and 
darkest parts above $45^{\circ}$. The brightest regions in field (b) 
have $\sim 8\,$MJy/sr surface brightness. The surface brightness falls 
to $5\,$MJy/sr in the regions where emission is still discernable and $1.5\,$MJy/sr in the darkest parts. 
} 
\label{fig:fig3}
\end{figure}

Fig.~\ref{fig:fig3} displays IRAS $100\,\mu$m images of regions (a) and (b).
Region (b) and the high latitudes areas ($b>45^\circ$) of region (a) each sample a low column density HLC.
The two lower Galactic latitude areas of field (a) are at the edge of the Scorpio-Centaurus association.  

The color of either Pioneer, ISGL, or DGL, is the ratio of the 
brightnesses in the R and B bands.
The larger it is, the redder the color under consideration.

\begin{figure*}
\resizebox{2\columnwidth }{!}{\includegraphics{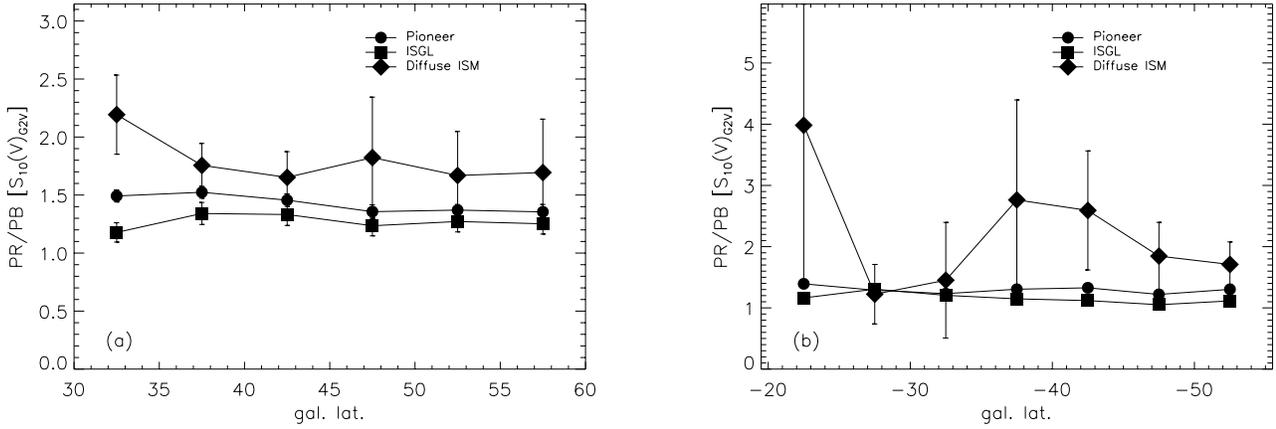}} 
\caption{GWF fig.~14. Caption: `Red/blue ratio 
for the Pioneer measurements, the ISGL, and the diffuse ISM. The 
first plot (a) displays the cut in Galactic longitude between  
$0^{\circ}$ and $ 5^{\circ} $ and the second plot (b), the cut 
between  $95^{\circ}$ and $ 100^{\circ}$.'} 
\label{fig:fig4}
\end{figure*}

The input data, from which all quantities will be deduced, of GWF's work are Pioneer's night sky brightnesses in the B and R 
bands (brightest stars apart), $P_{R}$ and $P_{B}$, on the one hand, the integrated starlight and 
galaxies brightnesses (brightest stars apart), $isgl_{R}$ and $isgl_{B}$, computed from 
several star catalogues, on the other.
Relative error margins estimated in GWF are $3\%$ and $2\% $ for $P_R$ and $P_B$ (sect.~2.1 in GWF), $5\%$ for $isgl_R$ and $isgl_B$.
Therefore, the error margins on Pioneer and ISGL colors are
\begin{eqnarray}
     \Delta \frac{P_R}{P_B}&\,=\, &
    \frac{P_R}{P_B}\left(
    \frac{\Delta P_R}{P_R}+\frac{\Delta P_B}{P_B}\right)\nonumber \\
    &\,\sim\, &0.08 \nonumber\\
    \Delta \frac{isgl_R}{isgl_B}&\,\sim\, &0.13
               \label{eq:err}
\end{eqnarray}
(with $P_R/P_B\sim isgl_R/isgl_B\sim 1.3$, Fig.~\ref{fig:fig4}).

The error on the color difference is $\sim 0.2$.
 \section{GWF demonstration of a non scattered component in DGL} 
 \label{gwfdem}
By subtracting the estimated ISGL to Pioneer B and R brightnesses,
GWF calculate $S_{B}$ and $S_{R}$, the B and R DGL brightnesses.
DGL color $S_{R}/S_{B}$ is then compared to ISGL's and Pioneer's.
GWF conclude:
``\emph{From Figure 14, it is obvious that the diffuse ISM 
is redder (larger $P_R/P_B$ ratio) than either the Pioneer measurements 
or the ISGL. As the scattered component of the diffuse ISM (DGL) is 
bluer (see \S~4.1) than Pioneer measurements, this requires that the 
nonscattered component is present in the red diffuse ISM intensity.}''
(GWF, p.531).

Fig.~14 of GWF is reproduced here (Fig.~\ref{fig:fig4}).

GWF demonstration of a non-scattered component in DGL is thus divided 
into  two.
GWF first attempt to show from observations that DGL is redder than ISGL 
and Pioneer.
In the mean time, \S~4.1 of GWF relies on the Witt-Petersohn 
(`WP') model \citep{witt94} to show that scattered-DGL should 
be bluer than ISGL and Pioneer.
The contradiction between observations and the WP model leads 
the authors to assume there is ERE in DGL.
 \section{Comparison of ISGL, DGL and Pioneer colors} \label{obs}
 \subsection{On the detection of ERE in DGL at $l=0^\circ$ and $l=100^\circ$} \label{ere}
The second part of the GWF demonstration will easily be accepted: HLCs, as long as they are of low column density (presumed to be the case at high Galactic latitudes), must give a bluer scattered-DGL than ISGL does. 
In view of GWF fig.~14 (here Fig.~\ref{fig:fig4}), and of the very large error bars on DGL color, the problem is rather to determine how precise is the comparison of DGL color to Pioneer's or ISGL's?

The comparison, undertaken in GWF, of DGL's to Pioneer's or ISGL's 
colors is extremely unprecised 
due to the considerable amplification of the error margin of DGL color 
(Fig.~\ref{fig:fig4}), a consequence of
the successive subtractions and division necessary to its 
estimate. 
However, as demonstrated below, this comparison can be done with much 
better accuracy through a direct comparison of the `input' data, 
ISGL and Pioneer colors.

Indeed, to compare $S_{R}/S_{B}$ and $isgl_{R}/isgl_{B}$ is like 
comparing $S_{R}isgl_{B}$ and $S_{B}isgl_{R}$.
This is also equivalent to the comparison of 
$S_{R}isgl_{B}+isgl_{R} isgl_{B}=isgl_{B}(S_{R}+isgl_{R})$ 
and of $S_{B}isgl_{R}+isgl_{R} isgl_{B}=isgl_{R}(S_{B}+isgl_{B})$; 
that is (Equalities~\ref{eq:p}) of
$isgl_{B}P_{R}$ and of $isgl_{R}P_{B}$.

Therefore, the comparison of $S_{R}/S_{B}$ and 
$isgl_{R}/isgl_{B}$ is strictly equivalent to the comparison of 
$P_{R}/P_{B}$ and of $isgl_{R}/isgl_{B}$.
One will similarly prove that the comparison of DGL color $S_{R}/S_{B}$ 
to Pioneer's $P_{R}/P_{B}$, is equivalent to that of ISGL and 
Pioneer.

We have
\begin{equation}
\frac{S_{R}}{S_{B}}> \frac{isgl_{R}}{isgl_{B}} \,
\Longleftrightarrow\,
\frac{P_{R}}{P_{B}}> \frac{isgl_{R}}{isgl_{B}}\,
\Longleftrightarrow\,
\frac{S_{R}}{S_{B}}>\frac{P_{R}}{P_{B}}
\label{eq:equi}
\end{equation}¥
Comparison of DGL and ISGL colors could have been resumed, with 
far better accuracy, to that of the input data, Pioneer's and ISGL's.

For DGL to be much redder than ISGL, Pioneer needs to be significantly redder than ISGL.
It is seen from the two bottom curves in each plot of  Fig.~\ref{fig:fig4} (fig.~14 in GWF) that this is generally not the case.
For all areas except the lower latitude one of field (a) (and maybe also the (a) area at $b=37^\circ$), ISGL and Pioneer colors are equal within the error margin.
It is therefore incorrect to deduce from this figure that observed DGL is redder than ISGL.

\begin{figure}
\resizebox{\columnwidth }{!}{\includegraphics{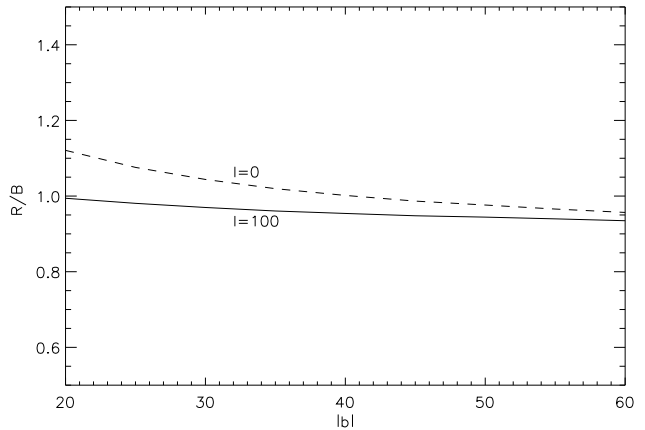}} 
\caption{R/B color of stars with mag $>6$ 
at $l=0$ and $l=100$. Computed from the Besan\c{c}on Galactic 
Model. A slight reddening of 0.6~mag per kpc is assumed by the model.} 
\label{fig:fig5}
\end{figure}

The only areas in which one could claim to have detected ERE are the two lowest latitudes of cut ($a$).
Notwithstanding, the ISGL $R/B$ color is abnormally low in these areas.
While it is expected to strongly increase when moving towards the Galactic plane  (see Fig.~\ref{fig:fig5}; the lowest latitude areas of field ($a$) cover regions with high column densities and should have an even redder ISGL color than shown on the figure), the GWF's ISGL color along the cut decreases at the lowest latitudes.
Therefore, either these two areas do not follow standard Galactic laws, or, the error margin given in GWF are underestimated (at least in these areas).
The later case would imply that it is not possible to guarantee that Pioneer is, in these areas, much redder than ISGL.
In the former, since $A_V$ is much higher in these areas than at higher latitudes (Sect.~\ref{lowdens}), it is still necessary to prove that scattered-DGL must be bluer than ISGL (this will be discussed in Sect.~\ref{sss}).
Dense clumps in MCLD~123.5~+24.9 for instance have nearly no blue surface brightness, and thus appear red, but this is an effect of extinction, not of ERE.
 \subsection{Error bars in the GWF data} \label{eb}
The too blue ISGL color found in field (a) low latitude  areas could result from the method used to calculate B and R magnitudes of stars in the GWF Master Catalog, generally  from the magnitude at one wavelength alone.
For stars and galaxies with a $P_B$ magnitude between 9.5 and 13, B and R magnitudes are extrapolated either from the V ($\lambda=5600\,\rm\AA$) or J ($\lambda=4500\,\rm\AA$) magnitudes of the Guide Star Catalog (sect.~3.2 in GWF).
For stars between 6.5 V magnitude and 9.5 $P_B$ magnitude, with no available B and/or R magnitudes, GWF will estimate the R and B magnitudes from the V one, with an algorithm which supposes a uniform reddening, dependent on the stars' distance only.
This algorithm is the same for all regions, and does not
take into account the much larger extinction in the lower latitude 
areas of field (a).
It would be good to know, in each area, what proportion of ISGL's B and R brightnesses was extrapolated by this method.
A possible correction would necessarily go in the sense of a redder ISGL (and a color closer to Pioneer's) than shown on the left plot of Fig.~\ref{fig:fig4} for the low latitude points.
In this case, the conservative $5\%$ error margin adopted in GWF (sect.~2.1) for the calculation of ISGL
is underestimated.

A second puzzling question on the GWF data concerns the ratios of DGL to ISGL in B and in R.
These ratios, are, from figs.~12 and 13 in GWF (the values are unfortunately not tabulated in the paper),  of order $40\%$ in $R$ and $30\% $ in B (corresponding to DGL/Pioneer ratios of $30\%$ and $25\%$).
This represents a considerable amount of DGL, even higher to what is found in the direction of  the Galactic plane ($\sim 30\%$).
From \citet{toller81}, or table~39 of \citet{leinert98}, $DGL/ISGL$ ratios should be less than $20\%$ for $|b|\geq 30^\circ$, with no pronounced dependence on Galactic longitude.
A too large DGL/ISGL ratio can result either from an over-estimate of Pioneer brightnesses, or, from an under-estimate of ISGL.
A correction would necessarily increase the proportion of ISGL in Pioneer, and here again, contribute to make Pioneer and ISGL colors on Fig.~\ref{fig:fig4} even more indistinguishable.
 \section{Modeling Pioneer data at high latitudes} \label{wp}
Beside the question of an emission process in the red,  the GWF paper raises the important problem of the pertinence of a model, and of the reliability of the conclusions one will deduce from it.
To show that scattered-DGL 
should be bluer than ISGL (or Pioneer), and to estimate the strength of 
the non-scattered DGL (due to emission), the authors compute from a model which is meant to 
represent the Galaxy (``\emph{Such a model exists and it is the 
Witt-Petersohn DGL model\ldots The WP model treats the Galaxy as a 
gigantic reflection nebula.}'' GWF, p.23) the color of scattered DGL.
The model of the Galaxy is constructed from HI measurements in different directions, from which the authors will deduce a structure for the interstellar medium on the lines of sight. 
 
The WP model predictions never match the R band observations of DGL.
For field (a) the predicted DGL scattered light surface brightness is 
lower than the 
observed one, which the authors  attribute to the presence of ERE 
(fig.~16 in GWF); 
in field (b) however, the predicted R brightness is curiously larger than the 
observed one (where is the ERE?).
In the blue (fig.~15 in GWF), where ERE is not suspected, 
there is no more agreement between the model and the observations:
the model fails to reproduce the shape of observed DGL in field (a);
it is larger for field (b) by a factor of 4.

On what basis is it possible to accredit the conclusions of a model which is so far from the reality of the observations it is meant to describe?
GWF argue that any change of parameters in the model will give the 
same scattered-DGL color, bluer than ISGL and Pioneer (fig.~17 of GWF), 
and therefore that observed DGL is definitely redder than it 
should be if it was scattering alone. 
This may be inherent to the GWF-WP model, and certainly true for scattering through a low column density medium (in which case no model is necessary), but does it mean that 
all models will give the same color? 
When will  DGL be  bluer than ISGL and Pioneer, 
and can't it be redder?

Models for the interstellar medium have been used (\citet{mattila78}, and references therein) to study the properties of DGL, and may be justified for regions close to the Galactic plane, which  sample sight-lines with separated layers of different and high optical depths.
The situation at high Galactic latitudes is much different.
In HLCs with low column densities on the average, in which CO is generally not detected,  high optical depth clumps must give a marginal contribution to the overall brightness of the cirrus at visible wavelengths, which should allow the use of low column densities approximations, and avoid constraining the structure of the interstellar medium in a way we are not able to verify.
This might provide an easier and more realistic way to understand the observations at hand.
 \section{Analysis of DGL data at high Galactic latitudes} \label{lowdens}
Field ($b$), and the areas of field ($a$) with a Galactic latitude of over $45^{\circ}$, have 100~$\mu$m surface brightnesses between 1.5~MJy/sr 
and 5~MJy/sr (the infrared surface brightness decrease with increasing Galactic latitude is due to the decrease of the radiation field).
A $100\,\mu$m to visible extinction ratio $I_{100}/A_V$ of 
$18$~MJy/sr/mag \citep{boulanger88}, will give $A_V$ values of less than 0.3~mag. 

The same order of magnitudes will be deduced from the NASA/IPAC ExtraGalactic Database (NED: http://nedwww.ipac.caltech.edu/index.html)  calculator of Galactic extinction \citep{schlegel98}.
NED provides, from the analysis of infrared IRAS and DIRBE images, mean values of $E(B-V)$ averaged over fields $\sim 6'$ wide, with a precision of $\sim 15\%$. 
For all but the two lowest latitude areas in (a), $A_V$ ranges from 0.1 to 0.3~mag., with corresponding $A_R$ between 0.07 and 0.25, and  $A_B$ between 0.1 and 0.4.
$A_V$, $A_B$, $A_R$ given by NED are deduced from $E(B-V)$ and standard conversion factors (Table~3 of \citet{cardelli89}).
For the two lowest latitudes of (a), NED will find $A_V$ between 0.5 and 0.7, $A_R$ between 0.3 and 0.6, and $A_B$ between 0.6 and 1.

The low average optical depth values found in all bands and in most of the GWF areas imply that high column density clumps, if there are any, occupy a very small fraction of the sky and do not contribute much to DGL in these directions.
Except, maybe, for the two lowest latitude areas in (a), the radiative transfer through the cirrus on the line of sight, independently of its structure, can be treated using the single scattering approximation \citep{zagury99} (eventually the low column density approximation when $A_B$ is sufficiently small).
These approximations constitute, for a given optical depth, an upper value of the scattered light surface brightness (in the front direction, it will diminish with multiple scattering).  

For one direction of the sky (one direction at high $|b|$, as considered by GWF), and a given optical band (B, R, or V), let $\tau$ be the average optical depth of the cirrus in that direction, and $\sigma$ the surface brightness (units of power per unit area per unit solid angle and unit wavelength) of direct starlight corrected for the cirrus reddening.

I introduce  $s$, the proportion of  $\sigma$  reddened by cirrus on the line of sight ($s\leq 1$).
 $(1-s)\sigma$ is the proportion of  $\sigma$ which either comes from stars in front of the cirrus, or from background stars seen through a hole.
$s$ is linked to the spatial repartition of the stars and to the filling factor of the cirrus.
 Blue stars being, on the average, closer to the Galactic plane $s_R$ should be larger than $s_B$.
$s$, as shown below, and will be an important parameter when the area which is considered contains directions with much different optical depths than the average.

$q$ will be the proportion of $s\sigma$ forward scattered by the cirrus, which  is equal to the albedo, $\omega$, of interstellar grains, if single scattering applies and for a large forward scattering phase function of the grains.  
The exact value of $\omega$, which is known to be large, is not a critical point for the orders of magnitudes which will be derived.
For calculation purposes I will use, as GWF did and as it is traditionally accepted (see fig.~5 of \citet{mathis77,draine03}; GWF and references therein): $q=\omega=\omega_R=\omega_B\sim 0.6$.

The surface brightness, $S$, of light forward scattered by a cirrus on the line of sight can be estimated from the energy balance between the forward directed light  impinging  on the cirrus, and coming out from it.
The light absorbed by the cirrus corresponds to a surface brightness, using the single scattering approximation, of $(1-\omega\tau)se^{-\tau}\sigma$.
Consequently
\begin{eqnarray}
 S &\,=\, &s e^{-\tau}\sigma-\left( \left(1-\omega\tau\right)se^{-\tau}\sigma\right) \nonumber\\
 &\,=\, & \omega\tau se^{-\tau}\sigma
 \label{eq:s}
\end{eqnarray}
$S$ does not depend on the exact shape of the phase function (as long as it is forward directed).

Then
\begin{eqnarray}
    isgl &\,=\, &\left(1-s+se^{-\tau}\right)\sigma 
    \nonumber \\
     S &\,=\, & \omega\tau se^{-\tau} \sigma
    \nonumber \\
     P &\,=\, &  \left( 1-s+se^{-\tau}+\omega\tau se^{-\tau}\right) \sigma  \nonumber\\
 \frac{isgl_R}{isgl_B} &\,=\, &
    \frac{ 1-s_R+s_Re^{-\tau_R}}{1-s_B+s_Be^{-\tau_B}}\frac{\sigma_R}{\sigma_B}
      \nonumber \\
  \frac{S_R}{S_B} &\,=\, &
    \frac{\tau_R}{\tau_B}e^{\tau_B-\tau_R}\frac{\sigma_R}{\sigma_B}
   \nonumber \\
 \frac{P_R}{P_B} &\,=\, &
    \frac{1-s_R+s_Re^{-\tau_R}+\omega\tau_R s_Re^{-\tau_R}}{1-s_B+s_Be^{-\tau_B}+\omega\tau_B s_Be^{-\tau_B}}\frac{\sigma_R}{\sigma_B}
      \label{eq:ssa}
\end{eqnarray}
Since  ISGL is observed to be redder than the source radiation field $\sigma$,
$(1-s_R+s_Re^{-\tau_R})/(1-s_B+s_Be^{-\tau_B})$ must be larger than 1.
With $\tau_B\sim 1.8\tau_R$, it implies that $s_B$ is larger than $0.5s_R$, and therefore: $s_R\geq s_B\geq 0.5s_R$.

In any band, the ratio of DGL to ISGL is  (Eqs.~\ref{eq:ssa})
\begin{eqnarray}
        \frac{S}{isgl}&\,=\, &\omega e^{-\tau}\frac{s\tau }{1-s\left(1-e^{-\tau}\right)}\nonumber\\
       & \,\sim \, &\omega e^{-\tau}\frac{s\tau }{1-s\tau}
               \label{eq:ssr}
\end{eqnarray}
The last approximation holds for the areas in which NED finds low column densities on the average (up to $\tau\sim 0.5$).
In these areas, for $S/isgl$  to be between 0.15 and 0.20  (table~39 of \citet{leinert98}), or more (GWF), with an albedo $\omega\sim 0.6$, $s$ needs to be close to 1.
Therefore most of the direct starlight  we receive from fields (a) and (b) comes from behind the cirrus on the lines of sight (Fig.~\ref{fig:fig3}) and is largely reddened by them.
With $s\sim 1$, from Eqs.~\ref{eq:ssa}
\begin{eqnarray}
    isgl &\,=\, & e^{-\tau} \sigma
    \nonumber \\
     S &\,=\, &\omega\tau e^{-\tau} \sigma
    \nonumber \\
     P &\,=\, & \left( 1+\omega\tau\right) e^{-\tau}\sigma  \nonumber\\
   \frac{S}{isgl}& \,=\, & \omega\tau  \nonumber \\
    \frac{isgl}{P} &\,=\, & \frac{1}{1+\omega\tau} 
       \label{eq:ssa1}
\end{eqnarray}
If  $S/isgl$ is less than $ 0.2$  at high latitudes (table~39 of \citet{leinert98}) $\tau$ is less than $0.3$, which is what NED finds over field (b) and the high latitudes areas of field (a).
Scattered-DGL to ISGL ratios larger than 0.3 in B, as it is found  in GWF (figs.~12 and 13),  would imply larger optical depths ($A_B\geq 0.5$) along these sight-lines.

The ratio of the input data ISGL and Pioneer (last equality of Eqs.~\ref{eq:ssa1}), $isgl/P$, should be larger in R than in B, with $(isgl/P)_R-(isgl/P)_B \sim 0.5\omega\tau_B$  in between $0\%$ and $\sim 10\%$.
 GWF find a larger ratio in  B ($ 70$ to $75\%$ against 65 to $70\%$ in R, fig.~12 in GWF), and a difference of $\sim -5\%$.
This  difference between expectations and the GWF data could be related to the possible sources of error on Pioneer and ISGL discussed in Sect.~\ref{eb}.

With $s\sim 1$, colors are
\begin{eqnarray}
    \frac{isgl_R}{isgl_B} &\,=\, &
    e^{\tau_B-\tau_R}\frac{\sigma_R}{\sigma_B} \sim e^{0.5\tau_B} \frac{\sigma_R}{\sigma_B}
      \nonumber \\
  \frac{S_R}{S_B} &\,=\, &
    \frac{\tau_R}{\tau_B}e^{\tau_B-\tau_R}\frac{\sigma_R}{\sigma_B} \sim 0.5e^{0.5\tau_B}\frac{\sigma_R}{\sigma_B}
   \nonumber \\
 \frac{P_R}{P_B} &\,=\, &
    \frac{1+\omega\tau_R}{1+\omega\tau_B}e^{\tau_B-\tau_R}\frac{\sigma_R}{\sigma_B}
    \sim \frac{1+0.5\omega\tau_B}{1+\omega\tau_B}e^{0.5\tau_B}\frac{\sigma_R}{\sigma_B}\nonumber\\
             \label{eq:col1}
\end{eqnarray}
The color difference between Pioneer and ISGL will be
\begin{eqnarray}
    \frac{P_R}{P_B}-\frac{isgl_R}{isgl_B} &\,=\, &
   -\omega    \frac{\tau_B-\tau_R}{1+\tau_B\omega} e^{\tau_B-\tau_R}\frac{\sigma_R}{\sigma_B}
      \nonumber \\
  &\,\sim\, &
    - \frac{0.5\omega A_B}{1+\omega A_B}e^{0.5A_B}\frac{\sigma_R}{\sigma_B}
             \label{eq:coldiff}
\end{eqnarray}
$A_B\sim 0.3$ will give a color difference of $0.1$ in absolute value, and a conservative $A_B=0.5$ of less than 0.17.
ISGL and Pioneer are thus expected to have very close colors, which agrees, despite the possible sources of error in the data  (Sect.~\ref{obs}),  with Fig.~\ref{fig:fig4}.
To establish the presence of ERE in DGL,  it is necessary to prove that Pioneer is redder than ISGL, and that the colors are separated with better precision than the present $0.2$ (Sect.~\ref{not}).

Eqs.~\ref{eq:col1} also show that, as long as one can consider a mean reddening over the field, DGL will be twice bluer (even bluer than found in GWF), than ISGL or Pioneer.
This may not always be the case however.
If the medium is structured, and contains clumps of high optical depth but with small surface coverage,  the color of ISGL could be that of the region outside the clumps and not differ much from the color of $\sigma$, while the color of DGL will be $\sim 0.5e^{0.7A_V}\sigma_R/\sigma_B$.
DGL will be redder than ISGL for $A_V$ values larger than $\sim 1$.

For high column density clumps (of optical depth $\tau$, large enough for blue light to be neglected, and with small overall surface coverage) immersed in a medium of negligible optical depth ($isgl\sim \sigma$), one will obtain
\begin{eqnarray}
     \frac{P_R}{P_B} &\,=\, &
  \frac{isgl_R+S_R}{isgl_B+S_B} \sim \frac{\sigma_R+S_R}{\sigma_B} \nonumber\\
  &\,\sim\, &\frac{\sigma_R}{\sigma_B}+ \frac{S_R}{\sigma_B} 
             \label{eq:clump1}
\end{eqnarray}
and
\begin{equation}
        \frac{P_R}{P_B}-\frac{isgl_R}{isgl_B} \,\sim\, \frac{S_R}{\sigma_B}\,\leq\, \omega\tau_R se^{-\tau_R}\frac{\sigma_R}{\sigma_B},
               \label{eq:clump2}
\end{equation}
since single scattering gives an upper limit of scattered light brightness.
Pioneer is now, as DGL, redder than ISGL.
But, while the difference of reddening between DGL and ISGL can be large, the color difference between Pioneer and ISGL will remain small, less than $0.2s$ (with $\omega\sim 0.6$ and $\sigma_R/\sigma_B \sim 1$).
\subsection{Discussion} \label{dis}
\subsubsection{Low column densities regions} \label{lc}
With little assumptions the preceding calculations give a quantitative account of observations of night sky background starlight at high Galactic latitudes.
It is clear from Eqs.~\ref{eq:col1} that in 
areas where column densities are not too high, DGL will be much bluer than the 
unreddened starlight color (by a factor of 2) and thus than ISGL or Pioneer.
However, Pioneer color in these areas 
is not determined by DGL, but by ISGL, and the color difference between ISGL and Pioneer is expected to be extremely small, as was observed.

These results are in conformity with what simple reasoning would give.
At high latitudes, extinction is e.g. low, Pioneer should be dominated by 
the ISGL, which suffers little extinction, while DGL represents 
a smaller fraction of Pioneer's surface brightness, because little 
extinction also means little scattering.
Thus, Pioneer's color should ressemble that of ISGL. 
On the other hand, DGL is expected to be much bluer, in the proportion of 
$\tau_{R}/\tau_{B}\sim 0.5$.
However, because DGL is weak, its color won't be easily determined, unless Pioneer and ISGL 
brightnesses can be set precisely.

The interpretation of GWF fig.~14 (here Fig.~\ref{fig:fig4}) for 
areas in field (a) at $b>45^{\circ}$, and over all field (b), is 
then straightforward:
in these areas, ISGL and Pioneer have very close colors which cannot be separated within the GWF error 
margin.

Local increases of column density in these areas, due to the small scale structure of 
the medium, will not modify the low column density approximation.
Because of the low optical depths NED finds in field (b) and for the (a)-field areas at
$b>45^{\circ}$, the number of such high density clumps must remain marginal.
Note that CO is not detected in these areas.
\subsubsection{Effect of an increase of column density on the observed 
colors} \label{sss}
To get better insight into the effect an increase of column density 
can have, I have considered (end of Sect.~\ref{lowdens}) the case in which one of the GWF areas would be composed of 
very dense clumps (dense enough to extinguish background starlight, 
and absorb all blue light, as for the dense clump mentioned in 
\citet{zagury99}) with an appreciable surface 
coverage,
embedded in a low column density medium.
The color of DGL, given by the color of the clumps, will be red.
The color of ISGL, on the contrary, will be given by the stars 
observed through the low column 
density medium, and be much bluer than DGL.
In this case, scattered-DGL, and Pioneer, can be redder than ISGL.

We see that the relative colors of DGL, Pioneer, and ISGL, depend on 
the column density and on the structure of the interstellar medium.
The claim that DGL should always be bluer than ISGL, whichever 
model is used, is therefore not correct.

The (a) areas under $b=45^{\circ}$ (3 data-points on 
Fig.~\ref{fig:fig4}) are in the densest parts of the HI loop at 
the edge of the Scorpio-Centaurus association and have high infrared surface 
brightnesses. 
Zeta Oph ($m_V=2.6$), a few degrees apart, may 
contribute to the 
heating of the region and to the enhancement, up to $30$~MJy/sr, of 
its $100\,\mu$m infrared emission. 
Notwithstanding, the column density must be higher in these areas than it 
is in the higher latitudes ones of field (a), or in field (b).
There is, in these areas, an increase 
in the average HI column density \citep{degeus88}, and hence of 
the visible extinction, attested by the much higher values found by NED;
CO is detected at the brightest IRAS positions \citep{laureijs95}.
The highest column densities are also proved by the drop 
in ISGL brightnesses (GWF, fig.~12, left plot) GWF finds in these areas.

It is improbable, however, that the higher column densities in these regions can explain the difference between Pioneer and ISGL colors observed on Fig.~\ref{fig:fig4}.
It would first require clumps with $A_V$ values larger than 1 and an appreciable surface coverage (end of Sect.~\ref{lowdens}).
Even so, the color difference between ISGL and Pioneer should be much less than 0.2 (if DGL is scattered light only).
As pointed out in Sect.~\ref{eb}, this difference needs to be confirmed by a more precise estimate of the uncertainties in the GWF data.
\section{Conclusion} \label{con}
This study of DGL at high $|b|$ has first unravelled several problems in the method and  data  GWF have used to propose detection of ERE in DGL.
It was also found that the near equality between Pioneer and ISGL colors, found by GWF, is expected if DGL results from the scattering of background starlight  by interstellar dust in the cirrus on the line of sight.
Also, orders of magnitudes have been calculated which agree with observation and do not introduce assumptions on the structure of the interstellar medium, for the strength and color of Pioneer, ISGL, and (scattered) DGL, at high Galactic latitudes.

Concerning the GWF paper, it was remarked that comparison of DGL and Pioneer or ISGL colors could be reduced, with far better accuracy and no need of estimating  the  color of DGL, to the comparison of Pioneer and ISGL colors.
This comparison shows that, in most areas, Pioneer and ISGL colors are equal within the error margins.
It is therefore not possible to conclude, in these areas, on the presence of ERE in DGL.

It is only in the two areas of field (a) at latitudes under $45^{\circ}$ that Pioneer and ISGL colors are separated enough to indicate a possible detection of ERE.
However, column densities in these areas, at the edge of Scorpio-Centaurus association, are much larger than at higher latitudes, so that there is no more guaranty that DGL should be bluer than ISGL.

Detection of ERE in GWF coincides with peculiarities of the GWF data which would need a detailed examination: a too blue color of ISGL in the low latitudes areas of field (a), and a much too large proportion of DGL in Pioneer, in all areas. 
The algorithm used to derive ISGL's B and R brightnesses, from the V magnitude of the stars, doesn't integrate the larger optical depths along some lines of sight, and may contribute to its too blue color in the lowest latitudes of field (a).
The  much larger  DGL/ISGL ratio ($\geq 30\,\%$) than expected ($<20\,\%$, \citet{toller81,leinert98}), especially in the red,  found in GWF, may indicate that Pioneer background night sky brightnesses are overestimated or that ISGL's are underestimated.
Correction of either one or the other effects will bring ISGL and Pioneer colors of GWF's fig.~14 even closer.
Therefore, either uncertainties have been underestimated, or, Galactic longitudes $l=0$ and $l=100$ have specific, remarkable properties which  need to be investigated. 

The investigations and orders of magnitude derived in Sect.~\ref{lowdens}, for Pioneer, ISGL, and scattered-DGL brightnesses and color, are consistent with observation.
They show the agreement between the observed range of average optical depths given by NED in fields (a) and (b), and the observed ratios of DGL to ISGL given by \citet{toller81} or table~39 of \citet{leinert98}, assuming all of DGL is scattered starlight.
They explain the close colors of ISGL and Pioneer, which cannot be separated within the precision of the observations, found in most of the GWF areas.
These investigations also prove that scattered-DGL can take a wide range of colors, from much bluer than ISGL and 
Pioneer in  not too high column density media, to much redder than 
ISGL if the medium is structured and contains clumps of
sufficiently high $A_{V}$ values.
 ISGL's color should always stay close to that of Pioneer.
 
Pioneer  data can therefore be well 
understood, without the use of a hypothetical model of the interstellar medium, within the traditional
comprehension of DGL, which is the scattering of background starlight by 
dust embedded in foreground interstellar cirrus.
In this respect, IRAS images, combined with reddening estimates and models of the
distribution of stars in the Galaxy, appear to be a powerful tool for the
analysis of DGL observations.

It will also be noted that this standard interpretation of DGL 
was found to hold well for the visible images 
of HLC MCLD~123.5~+24.9, at a resolution 
better than $2''$, in comparison to the $5^{\circ}\times 
5^{\circ}$ GWF boxes.

In view of the analysis here of the data presented by GWF, the detection of ERE in the diffuse interstellar medium seems premature, and requires further investigations.
\section*{Acknowledgments}
This work was partially funded by a NATO fellowship.

{}
\bsp

\label{lastpage}

\end{document}